\documentclass[10pt,aps,twocolumn,prc,nofootinbib,noshowkeys,superscriptaddress,floatfix]{revtex4-1}

\usepackage{amsmath,amsfonts,amsthm,amssymb}
\usepackage[dvips]{graphics,graphicx}
\usepackage[usenames,dvipsnames]{color}
\definecolor{darkblue}{RGB}{0,0,196}
\definecolor{darkgreen}{RGB}{0,120,0}
\usepackage[colorlinks=true,linktocpage=true,linkcolor=darkblue,citecolor=red,urlcolor=darkblue]{hyperref}
\usepackage{cancel}
\usepackage{multirow}
\usepackage{longtable}
\usepackage{color}
\usepackage[normalem]{ulem}
\usepackage{hyperref}
\usepackage{bigints}
\usepackage{xparse}
\usepackage{physics}
\usepackage{verbatim}
\usepackage{minibox}
\usepackage{comment}
\usepackage{appendix}

\begin{document}

\preprint{}

\title{Relaxation time approximation with pair production and annihilation processes}

\author{Samapan Bhadury}
\email{samapan.bhadury@niser.ac.in}
\affiliation{School of Physical Sciences, National Institute of Science Education and Research, HBNI, Jatni-752050, India}
\author{Wojciech Florkowski}
\email{wojciech.florkowski@uj.edu.pl}
\affiliation{Institute of Theoretical Physics, Jagiellonian University, PL-30-348 Krakow, Poland}
\author{Amaresh Jaiswal}
\email{a.jaiswal@niser.ac.in}
\affiliation{School of Physical Sciences, National Institute of Science Education and Research, HBNI, Jatni-752050, India}
\author{Radoslaw Ryblewski}
\email{radoslaw.ryblewski@ifj.edu.pl}
\affiliation{Institute of Nuclear Physics Polish Academy of Sciences, PL-31342 Krakow, Poland}

\date{\today}

\begin{abstract}
We extend the Boltzmann equation in the relaxation time approximation to  explicitly include transitions between particles forming an interacting mixture. Using the detailed balance condition as well as conditions of energy-momentum and current conservation, we show that only two independent relaxation time scales are allowed in such an interacting system. Dissipative hydrodynamic equations and the form of transport coefficients is subsequently derived for this case. We find that the shear and bulk viscosity coefficients, as well as the baryon charge conductivity are independent of the transition time scale. However, the bulk viscosity and conductivity coefficients that can be attributed to the individual components of the mixture depend on the transition time.  
\end{abstract}

\pacs{25.75.-q, 24.10.Nz, 47.75.+f}
	
\keywords{RTA}


\maketitle

%
\section{Introduction}
\label{introduction}

Quantum chromodynamics (QCD) is the fundamental theory of strong interactions. High energy heavy-ion collision experiments at the Relativistic Heavy Ion Collider (RHIC) at Brookhaven and the Large Hadron Collider (LHC) at CERN, Geneva, provide the opportunity to create hot and dense QCD matter and study its properties~\cite{Busza:2018rrf}. At very high energies, the quark and gluon degrees of freedom are liberated over the volume of colliding nuclei and produce the so called quark-gluon plasma (QGP). The phenomenological study of space-time evolution of QGP, by analyzing the experimental observables, helps us to understand its thermodynamic and transport properties \cite{Hwa:2004yg, Hwa:2010npa}. 

Relativistic dissipative hydrodynamics has been quite successful in explaining the experimental results indicating that QGP behaves like a nearly thermalized fluid (for recent reviews see, e.g., Refs.~\cite{Gale:2013da, Jeon:2015dfa, Jaiswal:2016hex, Romatschke:2017ejr, Berges:2020fwq}). Indeed the value of shear viscosity to entropy density ratio, extracted from hydrodynamic analysis of flow data (for recent results see Ref.~\cite{Bernhard:2019bmu}), was found  to be very close to the lower bound~\cite{Danielewicz:1984ww, Kovtun:2004de},  which led to the claim that QGP is the most perfect fluid ever observed.

Hydrodynamic modeling of relativistic heavy-ion collisions requires information about the microscopic dynamics of the system via equation of state and transport coefficients~\cite{Jaiswal:2016hex, Florkowski:2017olj}. This can be achieved by considering the dynamics of the microscopic degrees of freedom within the framework of relativistic kinetic theory \cite{Teaney:2013gca, Jaiswal:2013fc, Florkowski:2015lra, Jaiswal:2014isa, Denicol:2012cn, Jaiswal:2012qm, Ambrus:2017keg, Calzetta:2010au, Tsumura:2013uma, Tsumura:2015fxa, Liu:2019krs, Denicol:2010xn}. The quantities relevant for hydrodynamics can be obtained from relativistic kinetic theory by considering suitable moments of the phase-space distribution function \cite{Denicol:2010xn, Muller:1967zza, Israel:1979wp, Betz:2008me, Betz:2009zz, Martinez:2010sc, Martinez:2012tu, Florkowski:2013lza, Florkowski:2014bba, Bazow:2013ifa, Molnar:2016gwq, Molnar:2016vvu, Gao:2020vbh, Tinti:2015xwa, Dash:2019zwq}. Moreover, the space-time evolution of the phase-space distribution function is governed by the Boltzmann equation. Therefore, the hydrodynamic evolution of a system can be obtained from the moments of this equation. For a system close to equilibrium (which is believed to be true for strongly coupled QGP), the collision kernel of the Boltzmann equation can be simplified using the relaxation-time approximation (RTA)~\cite{PhysRev.94.511, Kumar:2017bja}. The RTA assumes that the effect of collisions is to exponentially drive the system towards local equilibrium~\cite{Baym:1984np}.

The relaxation-time approximation turns out to be very useful and has been employed extensively to derive the form of kinetic coefficients~\cite{ANDERSON1974466, Czyz:1986mr,Plumari:2012ep, Jaiswal:2014isa,Abhishek:2017pkp,Czajka:2017wdo, Mohanty:2018eja,Kurian:2018dbn} and dissipative hydrodynamic equations~\cite{Jaiswal:2013npa, Jaiswal:2013vta, Bhalerao:2013pza, Florkowski:2013lya, Florkowski:2013lza, Florkowski:2015lra,Bhadury:2020puc} as well as find their exact solutions \cite{Denicol:2014xca, Florkowski:2014sfa} and attractors~\cite{Denicol:2017lxn, Strickland:2018ayk, Jaiswal:2019cju, Strickland:2019hff, Dash:2020zqx}. While enormous simplification is achieved by considering RTA for the collision kernel, this comes at the expense of ignoring the interaction mechanism of the microscopic constituents. Moreover, RTA assumes a single timescale for thermalization of all types of microscopic interactions, whether they are elastic or inelastic. This may be a reasonable approximation if the timescale for inelastic processes is much smaller than elastic processes such that the chemical equilibration precedes the kinetic equilibration. On the other hand, if such separation of scales is not possible and the two timescales are comparable, RTA needs to be modified to correctly account for the elastic and inelastic processes separately (a first step in this direction, however, for a simplified space-time geometry, has been done in Refs.~\cite{Florkowski:2013uqa, Florkowski:2016qig}). 

In this article, we extend the Boltzmann equation in the relaxation time approximation to include explicitly the transitions between different particles forming an interacting  mixture. We consider two types of particles. The particles of the first type carry a conserved quantum number, hence, it is necessary to split them into particles and antiparticles. On the other hand, the number of particles of the second type is not conserved. In the following we call the particles of the first type ``quarks'' and of the second type ``gluons''. We also identify the conserved charge with the baryon number. We note, however, that these are to large extent symbolic names, since our particles may have several properties different from those characterizing quarks and gluons that appear in field-theoretic calculations. With this nomenclature remark in mind we may state that we consider inelastic interactions for the quark-antiquark annihilations to gluons and corresponding quark-pair production processes. 

Using the detailed balance condition as well as conditions of energy-momentum and current conservation, obtained from the Boltzmann equations, we show that there exists only two independent relaxation time scales in such an interacting system. Subsequently, we derive dissipative hydrodynamic equations for the evolution of this system as well as the corresponding transport coefficients.

The present article has the following structure: In Sec.~II we set up the necessary hydrodynamic framework required to study the space-time evolution of QGP using fundamental conservation laws. In Sec.~III we generalize the RTA to explicitly include inelastic interactions in the Boltzmann equation. Section IV deals with the derivation of the first-order transport coefficients and some limiting cases for the interacting medium are worked out therein. In Sec.~V we present numerical results for the transport coefficients and study effects coming from the presence of the transition time. Finally, in Sec. VI we summarize the key results of the present work and outline possible extensions of the present framework that can be studied in the close future.

\medskip
\textbf{Notations and Conventions:} In this article we use the following notations and conventions. The quantity $u^\mu$ is the fluid four-velocity (normalized to unity) and in the fluid rest frame $u^{\mu}=(1,0,0,0)$. The tensor $\Delta^{\mu\nu} = g^{\mu\nu} - u^\mu u^\nu$ is the projection operator that is orthogonal to the fluid velocity. The metric tensor is taken to be $g^{\mu\nu}=\text{diag}(1, -1, -1, -1)$. We choose the appropriate gluon and quark/antiquark degeneracy factors respectively as $g_g=N_s\times (N_c^2-1)$ and $g_q=N_s\times N_c\times N_f$, where $N_f=3$ is the number of flavors, $N_s=2$ is the spin degeneracy, and $N_c=3$ is the number of colors.


\section{Relativistic hydrodynamics}
\label{rel_hyd}

Hydrodynamic evolution of a relativistic system is governed by the conservation of energy-momentum tensor, $T^{\mu\nu}$, and particle four-current, $N^\mu$ (that we identify here, up to a factor of three, with the baryon current). In the present work we consider a system of interacting quarks, antiquarks and gluons. One can then write the conserved hydrodynamic quantities as moments of the phase-space distribution functions of quarks, antiquarks, and gluons, which can be further tensor decomposed in terms of the hydrodynamic degrees of freedom,
\begin{align}
T^{\mu\nu} &= \int dp\, p^\mu\, p^\nu \left[g_q \left( Q + {\bar Q} \right) + g_g\, G\right]\nonumber\\
&= \epsilon\,u^\mu u^\nu - (P + \Pi)\, \Delta^{\mu\nu} + \pi^{\mu\nu}, \label{Tmunu}\\
N^\mu &= \int dp\, p^\mu\, g_q\left(Q - {\bar Q}\right) = n\, u^\mu + n^\mu \label{Nmu},
\end{align}
where $dp \equiv  d^3p/[(2 \pi)^3\sqrt{{\bf p}^2+m^2}]$ is the Lorentz invariant momentum integral measure with $m$ being the particle mass, $p^\mu$ is the particle four-momentum, $g_q$, $g_g$ are the degeneracy factors for quarks and gluons respectively. In the above equations, $Q$, ${\bar Q}$ and $G$ are used to denote the distribution functions for quarks, antiquarks and gluons, respectively. The hydrodynamic variables $\epsilon$, $P$, and $n$ in the above equations are the energy density, pressure, and net number density of the system, respectively. The dissipative quantities, $\Pi$, $\pi^{\mu\nu}$, and $n^\mu$ are the bulk viscous pressure, shear stress tensor, and the dissipation current, respectively. In order to express $T^{\mu\nu}$ and $N^\mu$ in terms of the hydrodynamic variables, we define the fluid four-velocity in the Landau frame: $T^{\mu\nu}u_\nu=\epsilon u^\mu$. To define thermodynamic quantities for a non-equilibrium system, we have used the matching condition $\epsilon=\epsilon_{\rm eq}$ and $n=n_{\rm eq}$, where $\epsilon_{\rm eq}$ and $n_{\rm eq}$ are the corresponding equilibrium values.

The equilibrium quantities appearing in Eqs.~\eqref{Tmunu} and~\eqref{Nmu} can be written in terms of the equilibrium distribution functions as,
\begin{align}
\epsilon_{\rm eq} &= u_\mu u_\nu \int dp\, p^\mu\, p^\nu \left[ g_q \left(Q_{\rm eq} + {\bar Q}_{\rm eq}\right) + g_g\, G_{\rm eq} \right],\label{energy_density}\\
P_{\rm eq} &= -\frac{\Delta_{\mu\nu}}{3} \!\!\int\! dp\, p^\mu\, p^\nu \left[ g_q \left(Q_{\rm eq} + {\bar Q}_{\rm eq}\right) + g_g\, G_{\rm eq} \right],\label{pressure}\\
n_{\rm eq} &= u_\mu \int dp\, p^\mu\, g_q \left(Q_{\rm eq} - {\bar Q}_{\rm eq}\right)\label{number_density},
\end{align}
where $Q_{\rm eq}$, ${\bar Q}_{\rm eq}$, and $G_{\rm eq}$ are the equilibrium distribution functions for quarks, antiquarks, and gluons, respectively. In order to obtain the above relations, we use the identities $u_\mu u_\nu \pi ^{\mu\nu} = u_\mu n^\mu = 0$. 

Furthermore, we consider here the J\"uttner form of the classical Maxwell-Boltzmann distribution function for the equilibrium case. Since quarks and antiquarks have non-zero baryon chemical potential whereas gluons have zero baryon chemical potential, their equilibrium phase-space distributions are given by the expressions:
\begin{align}
Q_{\rm eq} &= e^{-\beta(u\cdot p) + \alpha}, \label{Q_eq}\\
{\bar Q}_{\rm eq} &= e^{-\beta(u\cdot p) - \alpha}, \label{Qbar_eq}\\
G_{\rm eq} &= e^{-\beta(u\cdot p)}, \label{G_eq}
\end{align}
where $\beta\equiv 1/T$ is the inverse temperature, $\alpha\equiv\mu/T$ is the ratio of baryon chemical potential and temperature, and $A\cdot B\equiv A_\mu B^\mu$. The above distribution is locally isotropic in momentum space. A mixture of quarks and gluons has also been studied previously with anisotropic distribution function in Ref.~\cite{Florkowski:2015cba}.

We note that the use of the classical distributions \eqref{Q_eq}--\eqref{G_eq} is appropriate for sufficiently dilute systems of particles. This implies that the parameter $\alpha$ cannot be too large. A natural range for $\alpha$ is given by the condition $|\alpha| < 4$, since for larger values of $|\alpha|$  the entropy density of the system described by Eqs.~\eqref{Q_eq}--\eqref{G_eq} may become negative (see, for example, Table 8.2 in Ref.~\cite{Florkowski:2010zz}). Thus, we will use the condition $|\alpha| < 4$ in our numerical calculations presented below in Sec.~V.

For a system close to equilibrium, one can write the non-equilibrium distribution functions as:
\begin{align}
Q(x,p) & = Q_{\rm eq}(x,p) + \delta Q(x,p), \label{Q_neq}\\
{\bar Q}(x,p) & = {\bar Q}_{\rm eq}(x,p) + \delta{\bar Q}(x,p), \label{Qbar_neq}\\
G(x,p) & = G_{\rm eq}(x,p) + \delta G(x,p), \label{G_neq}
\end{align}
where, $\delta Q$, $\delta{\bar Q}$, and $\delta G$ are the deviations from the equilibrium distribution functions of quarks, antiquarks, and gluons, respectively, satisfying $\delta Q \ll Q_{\rm eq}$, $\delta{\bar Q} \ll {\bar Q}_{\rm eq}$, and $\delta G \ll G_{\rm eq}$. The dissipative quantities appearing in Eqs.~\eqref{Tmunu} and \eqref{Nmu} can be expressed in terms of these deviations as
\begin{align}
\pi^{\mu\nu} &= \Delta^{\mu\nu}_{\alpha\beta} \int dp\, p^\alpha\, p^\beta \left[ g_q \left(\delta Q + \delta {\bar Q}\right) + g_g\, \delta G \right], \label{pimunu_KT}\\
\Pi &= -\frac{\Delta_{\alpha\beta}}{3} \!\!\int\! dp\, p^\alpha\, p^\beta \left[ g_q\! \left(\delta Q + \delta {\bar Q}\right)\! + g_g\, \delta G \right]\!, \label{Pi_KT}\\
n^{\mu} &= \Delta^{\mu}_{\alpha} \int dp\, p^\alpha\, g_q \left( \delta Q - \delta {\bar Q} \right). \label{nmu_KT}
\end{align}
The above expressions will be used later to derive the form of the dissipative equations and calculate the corresponding transport coefficients.

In covariant form, the relativistic hydrodynamic equations are given by vanishing four-divergence of energy-momentum tensor and conserved four-current, i.e., $\partial_\mu T^{\mu\nu}=0$ and $\partial_\mu N^\mu=0$. Using the second equalities in Eqs.~\eqref{Tmunu} and \eqref{Nmu}, and making appropriate projections we obtain
\begin{align}
\dot{\epsilon} + (\epsilon + P + \Pi) \theta - \pi^{\mu\nu} \sigma_{\mu\nu} &= 0,\label{continuity}\\
(\epsilon + P + \Pi) \dot{u}^\alpha - \nabla^\alpha (P + \Pi) + \Delta^\alpha_\nu \partial_\mu \pi^{\mu\nu} &= 0,\label{euler}\\
\dot{n} + n \theta + \partial_\mu n^\mu &= 0,\label{diffusion}
\end{align}
where $\theta \equiv \partial_\mu u^\mu$ is the expansion scalar, $\sigma^{\mu\nu} \equiv \Delta^{\mu\nu}_{\alpha\beta} (\nabla^\alpha u^\beta)$ is the shear tensor, $\dot{A} \equiv u^\mu \partial_\mu A$ is the co-moving derivative, $\nabla^\alpha \equiv \Delta^{\alpha\mu} \partial_\mu$ is the space-like derivative, and $\Delta^{\mu\nu}_{\alpha\beta} \equiv (\Delta^\mu_\alpha \Delta^\nu_\beta + \Delta^\mu_\beta \Delta^\nu_\alpha)/2 - (1/3) \Delta^{\mu\nu}\Delta_{\alpha\beta}$ is a 4-rank trace-less symmetric projection operator orthogonal to both $u_\mu$ and $\Delta_{\mu\nu}$. Equivalently, in terms of the distribution functions, the energy-momentum conservation and current conservation imply
\begin{align}
\int dp \, p^\mu\, p^\nu \partial_\mu \left[ g_q (Q + {\bar Q}) + g_g G \right] &= 0, \label{Tmunu_cons}\\
\int \, dp \, p^\mu \, \partial_\mu \left[ g_q (Q - {\bar Q}) \right] &= 0. \label{Nmu_cons}
\end{align}
From the above equations, we conclude that the space-time evolution of the distribution functions is necessary to determine the evolution of the system. In the following, we set up the Boltzmann equation within the kinetic theory framework to determine the space-time evolution of the phase-space distribution functions.


\section{Kinetic theory setup}
\label{kin_th}

For a dilute system, the evolution of a single particle phase-space distribution function, governed by the Boltzmann equation, is sufficient to characterize the system on a microscopic level. For a relativistic system, the Boltzmann equation in the relaxation-time approximation can be written as~\cite{ANDERSON1974466}
\begin{equation}\label{Boltz_eq_RTA}
p^\mu \partial_\mu f = -(u\!\cdot\!p)\,\frac{\delta f}{\tau_{\rm eq}}\,,
\end{equation}
where $f\equiv f(x,p)$ is the phase-space distribution function, $\delta f\equiv f-f_{\rm eq}$ is the deviation of the distribution function from equilibrium with $f_{\rm eq}$ being the equilibrium distribution function and $\tau_{\rm eq}$ is the relaxation time-scale in which a system approaches equilibrium. As discussed earlier, a single timescale of the above form may not be adequate to account for elastic as well as inelastic interactions, especially when these two scales are not sufficiently apart.

In order to account for inelastic interactions, we formulate an improvement of the Boltzmann equation with RTA collision term. For a system of quarks and gluons, we consider the inelastic processes $g\leftrightarrow q\bar{q}$. For such a system, we propose the following set of RTA Boltzmann equations,
\begin{align}
p^\mu \partial_\mu Q &= -(u\!\cdot\! p)\!\left[\, \frac{\delta Q}{\tau^q_{\rm eq}} - \frac{1}{2}\!\left( \frac{G}{\tau^{g\rightarrow q}_{\rm tr}} - \frac{Q + {\bar Q}}{\tau^{q\rightarrow g}_{\rm tr}} \right) \right]\!, \label{Beq_Q}\\
p^\mu \partial_\mu {\bar Q} &= -(u\!\cdot\! p)\!\left[\, \frac{\delta{\bar Q}}{\tau^{\bar q}_{\rm eq}} - \frac{1}{2}\!\left( \frac{G}{\tau^{g\rightarrow q}_{\rm tr}} - \frac{Q + {\bar Q}}{\tau^{q\rightarrow g}_{\rm tr}} \right) \right]\!, \label{Beq_Qbar}\\
p^\mu \partial_\mu G &= -(u\!\cdot\! p)\!\left[\, \frac{\delta G}{\tau^g_{\rm eq}} + r\!\left( \frac{G}{\tau^{g\rightarrow q}_{\rm tr}} - \frac{Q + {\bar Q}}{\tau^{q\rightarrow g}_{\rm tr}} \right) \right]\!,\label{Beq_G}
\end{align}
where $\tau^q_{\rm eq}$, $\tau^{\bar q}_{\rm eq}$, $\tau^g_{\rm eq}$ represent the relaxation time-scales for quarks, antiquarks, and gluons, respectively; $\tau^{q\rightarrow g}_{\rm tr}$, $\tau^{g\rightarrow q}_{\rm tr}$ represent the relaxation times for the processes $q\bar{q}\rightarrow g$ and $g\rightarrow q\bar{q}$, respectively, and
\begin{equation}
r = \frac{g_q}{g_g}. 
\end{equation}
It is important to note that the relaxation times, $\tau^q_{\rm eq}$, $\tau^{\bar q}_{\rm eq}$, $\tau^g_{\rm eq}$, $\tau^{q\rightarrow g}_{\rm tr}$ and $\tau^{g\rightarrow q}_{\rm tr}$ can, in general, be momentum dependent. However in the present study, which is a first step towards development of such an RTA framework, we treat these relaxation times to be momentum independent. We also note that in the usual relaxation-time approach, Eq.~\eqref{Boltz_eq_RTA}, only the first terms on the right-hand-side of Eqs.~\eqref{Beq_Q}--\eqref{Beq_G} are present. The additional inelastic-interaction terms are written keeping in mind that the process $q\bar{q}\rightarrow g$ increases the gluon distribution and decreases the distribution of quarks and antiquarks. Similarly, the reverse process $g\rightarrow q\bar{q}$ decreases the gluon distribution and increases the distribution of quarks and antiquarks. The factors $1/2$ and $r$ are introduced to ensure energy-momentum and net baryon current conservation, Eqs.~\eqref{Tmunu_cons} and \eqref{Nmu_cons}, as explained in more detail below.

At this juncture, we would like to emphasize that the usual RTA Boltzmann equation, Eq.~\eqref{Boltz_eq_RTA}, is quite general in the sense that one need not go into the mechanism of interaction. Rather, the relaxation time approximation states that the ``effect" of the interactions is to drive the system towards equilibrium, exponentially, with a time scale which is set by the relaxation time $\tau_{\rm eq}$. Let us consider the case in Eqs.~\eqref{Beq_Q}--\eqref{Beq_G} when all $\tau_{tr}\to\infty$. In this case we are left with the usual RTA Boltzmann equations for quarks, anti-quarks and gluons. It is important to note that this set of Boltzmann equations is not specific to any reaction/processes. This is apparent when one considers no quarks at the beginning of the evolution. However, due to the collision term, quarks are generated immediately. This is due to the fact that the presence of the rest of constituents leads to a thermal medium which in turn fixes the thermal distribution for quarks to be non-zero. We emphasize that in this sense, the RTA implicitly considers all interactions that lead to a thermalized state described by equilibrium distributions via the first terms on the right-hand sides of Eqs.~\eqref{Beq_Q}--\eqref{Beq_G}. A new feature of the present approach is that in addition to such general thermalization processes we explicitly include transitions between components of the system. The latter processes, by themselves, may not lead to equilibration and are constrained only by the conservation laws as shown later. The aim of the current work is to investigate whether such processes ``couple" to the thermalization processes and affect the values of the kinetic coefficients.

From Eqs.~\eqref{Beq_Q}--\eqref{Beq_G}, it might seem that there are five independent relaxation times, which characterize the timescales of equilibration of different collision processes. However, only two of them are truly independent as will be demonstrated in the following. We first note that as a consequence of the Landau frame and matching conditions, $u_\mu\delta T^{\mu\nu}=0$ and $u_\mu\delta N^{\mu}=0$, where $\delta T^{\mu\nu}\equiv T^{\mu\nu}-T^{\mu\nu}_{\rm eq}$ and $\delta N^{\mu}\equiv N^{\mu}-N^{\mu}_{\rm eq}$, with the equilibrium energy-momentum tensor and net baryon current being evaluated using the corresponding equilibrium distribution functions. Keeping this in mind, we see that in order to satisfy net baryon current conservation, Eq.~\eqref{Nmu_cons}, the required condition is $\tau^q_{\rm eq}=\tau^{\bar q}_{\rm eq}$. Similarly, in order to satisfy the total energy momentum conservation, Eq.~\eqref{Tmunu_cons}, we find the necessary condition to be $\tau^q_{\rm eq}=\tau^{\bar q}_{\rm eq}=\tau^g_{\rm eq}$. Therefore, within our framework, there is only one independent equilibration timescale possible, henceforth denoted as $\tau_{\rm eq}$ and corresponding to the three processes given in Eqs.~\eqref{Beq_Q}--\eqref{Beq_G}.

In order to find the relation between the two transition timescales, $\tau^{q\rightarrow g}_{\rm tr}$ and $\tau^{g\rightarrow q}_{\rm tr}$, we first rewrite the coupled Boltzmann equations, Eqs.~\eqref{Beq_Q}--\eqref{Beq_G}, in a compact form using matrix notation. Introducing vector notation for the distribution functions, namely
\begin{align}
\textbf{F}(x,p) &=
\begin{bmatrix}
\,Q(x,p)\,\\
{\bar Q}(x,p)\\
G(x,p)
\end{bmatrix}, \label{Fvec}\\
\textbf{F}_{\text{eq}}(x,p) &=
\begin{bmatrix}
\,Q_{\text{eq}}(x,p)\,\\
{\bar Q}_{\text{eq}}(x,p)\\
G_{\text{eq}}(x,p)
\end{bmatrix}, \label{Feqvec}
\end{align}
we rewrite Eqs.~\eqref{Beq_Q}--\eqref{Beq_G} as
\begin{align}\label{Beq_F}
p^\mu\partial_\mu \textbf{F} = -(u\!\cdot\! p)\Big[\hat{R}_{\rm eq}(\textbf{F}-\textbf{F}_{\rm eq}) + \hat{R}_{\rm tr}\textbf{F}\Big].
\end{align}
In the above equation, $\hat{R}_{\rm eq}$ and $\hat{R}_{\rm tr}$ are $3\times3$ square matrices, whose elements are inverse of the various relaxation time scales and are given by
\begin{eqnarray}
\hat{R}_{\rm eq} &=& \nu_{\rm eq}\!
\begin{bmatrix}
\,1\, & \,0\, & \,0\, \\
0 & 1 & 0\\
0 & 0 & 1
\end{bmatrix}\!, \label{Req}
\end{eqnarray}

\begin{eqnarray}
\hat{R}_{\rm tr} &=& \frac{1}{2}\!
\begin{bmatrix}
\nu^{qg}_{\rm tr} & \nu^{qg}_{\rm tr} & -  \nu^{gq}_{\rm tr}\\
\nu^{qg}_{\rm tr} &  \nu^{qg}_{\rm tr} & -  \nu^{gq}_{\rm tr}\\
\,-2 r \nu^{qg}_{\rm tr}\, & \,- 2 r \nu^{qg}_{\rm tr}\, & \,2 r \nu^{gq}_{\rm tr}\,
\end{bmatrix}\!, \label{Rtr}
\end{eqnarray}
where, we have introduced
\begin{align}\label{tau_nu}
\nu_{\rm eq} &= \frac{1}{\tau_{\rm eq}}\,, \quad \nu^{qg}_{\rm tr} = \frac{1}{\tau^{q\rightarrow g}_{\rm tr}}\,,  \quad \nu^{gq}_{\rm tr} = \frac{1}{\tau^{g\rightarrow q}_{\rm tr}},
\end{align}
to simplify our notation.

Now we will examine some constraints that are obtained from the fact that in global and local equilibrium cases, the Boltzmann equation gets simplified. In global equilibrium, the distribution function $\textbf{F} = \textbf{F}_0$ becomes constant and therefore its derivatives must vanish, i.e., $\partial_\mu \textbf{F}_0 = 0$. Note that the phase-space distribution function in global equilibrium is a solution of the Boltzmann equation. When the system approaches global equilibrium, the local equilibrium distribution function also attains a constant value, i.e., $\textbf{F}_{\rm eq} = \textbf{F}_0$. Hence in global equilibrium, the first term on right-hand-side of Eq.~\eqref{Beq_F} is zero and thus leaves us with the condition, $\hat{R}_{\rm tr}\, \textbf{F}_0 = 0$. Using this condition and assuming that the quarks and gluons have equal masses, we get the constraint:
\begin{align}\label{tau_tr_GEQ}
\tau^{q\rightarrow g}_{\rm tr} = 2\tau^{g\rightarrow q}_{\rm tr}\,\cosh{\alpha_0},
\end{align}
where the subscript `$0$' denotes the constant value of $\alpha$ in global equilibrium. Therefore the constraint obtained from global equilibrium condition reduces the number of independent relaxation times to two. It is important to note that if the quark and gluons masses are different, the transition relaxation time becomes momentum dependent. Herein, we restrict our considerations to the case where all masses are the same. 

Similarly, in local equilibrium the distribution function takes the form $\textbf{F} = \textbf{F}_{\rm eq}$, which is a function of space-time coordinates. It is important to note that the local equilibrium is defined as the maximum entropy state and, in general, it is not a solution of the Boltzmann equation. However, the condition that the distribution function has to be positive definite at all space-time points during its evolution requires that $\hat{R}_{\rm tr} \textbf{F}_{\rm eq} = 0$. Again, assuming that the quarks and gluons have equal masses, this yields a constraint between the two transition relaxation times,
\begin{align}\label{tau_tr_LEQ}
\tau^{q\rightarrow g}_{\rm tr} = 2\tau^{g\rightarrow q}_{\rm tr}\,\cosh{\alpha}.
\end{align}
Interestingly, the form of the constraint is similar to that obtained in Eq.~\eqref{tau_tr_GEQ} from global equilibrium consideration. It turns out that the above constraint, obtained using the condition of positive definiteness of the distribution function, can also be derived from entropy arguments as demonstrated below.

The entropy four-current for a mixture of quark, antiquark and gluon with identical particle masses can be written as \cite{DeGroot:1980dk}
\begin{equation}\label{entr}
S^\mu = - \sum_{k=1}^3 g_k\! \int\! dp \, p^\mu F_k \left( \ln F_k - 1 \right),
\end{equation}
where, $k=1,\,2,\,3$ represents quark, antiquark and gluon, respectively, with $g_k$ being the corresponding degeneracy factors and $F_k$ being the components of $\textbf{F}(x,p)$ defined in Eq.~\eqref{Fvec}. The four-divergence of $S^\mu$ in the above equation can be obtained as 
\begin{equation}\label{entr_div}
\partial_\mu S^\mu = -\!\int\! dp \, \textbf{M} \bullet \left( p^\mu \partial_\mu \textbf{F} \right) ,
\end{equation}
where, $\textbf{M}$ has components $M_k=g_k\ln F_k$ and $\textbf{A}\bullet\textbf{B}\equiv\sum_k A_k B_k $. Using Eq.~\eqref{Beq_F}, one can write the above equation as
\begin{equation}\label{entr_div_F}
\partial_\mu S^\mu = \!\int\! dp \, (u\!\cdot\! p)\, \textbf{M}\bullet \Big[\hat{R}_{\rm eq}(\textbf{F}-\textbf{F}_{\rm eq}) + \hat{R}_{\rm tr}\textbf{F}\Big].
\end{equation}
In local equilibrium, $\textbf{F}=\textbf{F}_{\rm eq}$, the divergence of entropy four-current should vanish, i.e., $\partial_\mu S^\mu_{\rm eq}=0$. From the above equation, we see that this condition can be fulfilled, with correct limiting approach to global equilibrium, if $\hat{R}_{\rm tr} \textbf{F}_{\rm eq} = 0$ which again leads to Eq.~\eqref{tau_tr_LEQ}.

Using the above constraint relation, the two transition timescales can be written in a parametric form in terms of a single transition relaxation time, $\tau_{\rm tr}$, as follows
\begin{equation} \label{tautr_param}
\tau^{q\rightarrow g}_{\rm tr} = \tau_{\rm tr}\,\sinh{2\alpha}, \quad
\tau^{g\rightarrow q}_{\rm tr} = \tau_{\rm tr}\,\sinh{\alpha}.
\end{equation}
Along with the previously obtained constraints from energy-momentum and net baryon current conservation, $\tau^q_{\rm eq}=\tau^{\bar q}_{\rm eq}=\tau^g_{\rm eq}=\tau_{\rm eq}$, the above paramterization leads to two independent relaxation time-scales: $\tau_{\rm eq}$ and $\tau_{\rm tr}$. We note that we assume that all the transition times are positive, hence, Eq.~\eqref{tautr_param} can be used if $\alpha > 0$. In the case where $\alpha < 0$, one should change $\alpha$ to $-\alpha$ in Eq.~\eqref{tautr_param} and follow in exactly the same way as in the case $\alpha > 0$. As it does not lead to any relevant differences, we may restrict our considerations to the case $\alpha >0$.

The set of coupled Boltzmann equations introduced by Eqs.~\eqref{Beq_Q}-\eqref{Beq_G} can now be re-expressed in terms of the independent relaxation times as
\begin{align}
p^\mu \partial_\mu Q &= -(u\!\cdot\! p)\!\left[\, \frac{\delta Q}{\tau_{\rm eq}} - \frac{1}{2\tau_{\rm tr}}\!\!\left(\! \frac{\delta G}{\sinh{\alpha}} - \frac{\delta Q + \delta {\bar Q}}{\sinh{2\alpha}} \!\right) \!\right]\!, \label{Beq_Q_fin}\\
p^\mu \partial_\mu {\bar Q} &= -(u\!\cdot\! p)\!\left[\, \frac{\delta{\bar Q}}{\tau_{\rm eq}} - \frac{1}{2\tau_{\rm tr}}\!\!\left(\! \frac{\delta G}{\sinh{\alpha}} - \frac{\delta Q + \delta {\bar Q}}{\sinh{2\alpha}} \!\right) \!\right]\!, \label{Beq_Qbar_fin}\\
p^\mu \partial_\mu G &= -(u\!\cdot\! p)\!\left[\, \frac{\delta G}{\tau_{\rm eq}} + \frac{r}{\tau_{\rm tr}}\!\left( \frac{\delta G}{\sinh{\alpha}} - \frac{\delta Q + \delta {\bar Q}}{\sinh{2\alpha}} \right) \right]\!,\label{Beq_G_fin}
\end{align}
where we have also used the fact that the right-hand-side of Eqs.~\eqref{Beq_Q}-\eqref{Beq_G} vanishes in equilibrium, i.e., arguments leading to Eq.~\eqref{tau_tr_LEQ}. The above equations are one of the main results of the present work and represent a generalization of the Boltzmann equation to explicitly include inelastic scattering in the relaxation-time approximation. We note here that the pair production mechanism within kinetic theory has also been studied in Refs.~\cite{Banerjee:1989by, Bhalerao:1997zc} by deriving a source term.

Before we turn to a systematic analysis of the near equilibrium behavior of the system described by Eqs.~\eqref{Beq_Q_fin}--\eqref{Beq_G_fin}, it is interesting to observe that there are two combinations of the distribution functions that may be interpreted as independent ones,
\begin{eqnarray}
Q^- &=& Q - {\bar Q}, \label{eq:Qminus} \\
S^+ &=& r (Q + {\bar Q}) + G = r Q^+ + G.
\label{eq:Splus}
\end{eqnarray}
By making appropriate linear combinations of Eqs.~\eqref{Beq_Q_fin}--\eqref{Beq_G_fin} one finds that both $Q^-$ and $S^+$ satisfy the same kinetic equation,
\begin{eqnarray}
p^\mu \partial_\mu Q^- &=&  -(u\!\cdot\! p) \frac{\delta Q^-}{\tau_{\rm eq}},
\label{eq:Qminuseq} \\
p^\mu \partial_\mu S^+ &=& -(u\!\cdot\! p) \frac{\delta S^+}{\tau_{\rm eq}}.
\label{eq:Spluseq}
\end{eqnarray}
It is interesting to notice that the evolution of $Q^-$ and $S^+$ is not affected by the transition processes. This is due to the detailed balance constraint included in Eqs.~\eqref{Beq_Q_fin}--\eqref{Beq_G_fin}. Clearly, Eqs.~\eqref{eq:Qminuseq} and \eqref{eq:Spluseq} should be supplemented by a third equation for yet another linear combination of the original distribution functions. 

For the third combination, it is convenient to use 
\begin{equation}
S^- = Q + {\bar Q} -  G = Q^+ - G,
\label{eq:Sminus}
\end{equation}
which satisfies the equation
\begin{eqnarray}
p^\mu \partial_\mu S^- &=& -(u\!\cdot\! p)
\left[ \frac{\delta S^-}{\tau_{\rm eq}}\!-\!
\frac{1\!+\!r}{\tau_{\rm tr}}
\left( \frac{G}{\sinh \alpha} - \frac{Q^+}{\sinh 2\alpha} \right) \right]. \nonumber \\
\label{eq:Sminuseq}
\end{eqnarray}
Using the fact that the right-hand side of Eq.~\eqref{eq:Sminuseq} vanishes for local equilibrium, we can rewrite it as
\begin{eqnarray}
p^\mu \partial_\mu S^- &=& -(u\!\cdot\! p)
\left[ \frac{\delta S^-}{\tau^{\rm eff}_{\rm eq}}\!+\!
\frac{\delta S^+}{\tau_{\rm tr}}
\left( \frac{1}{\sinh 2 \alpha} - \frac{1}{\sinh \alpha} \right) \right]. \nonumber \\
\label{eq:Sminuseq1}
\end{eqnarray}
where we have introduced an effective relaxation time for the distribution $S^-$
\begin{eqnarray}
\tau^{\rm eff}_{\rm eq} &=& \tau_{\rm eq} \,
\frac{\gamma}{\gamma + \frac{r}{\sinh\alpha} + \frac{1}{\sinh 2\alpha}}
\label{eq:taueff}
\end{eqnarray}
and the ratio 
\begin{equation}
\gamma = \frac{\tau_{\rm tr}}{\tau _{\rm eq}}.
\end{equation}
From Eq.~\eqref{eq:taueff} we conclude that the effective relaxation time is always smaller than the original relaxation time, $\tau^{\rm eff}_{\rm eq} \leq \tau_{\rm eq}$, and $\tau^{\rm eff}_{\rm eq} \to 0$ for $\gamma \to 0$ or $\alpha \to 0$, i.e., for very small transition times. 


\section{Transport Coefficients}
\label{tr_coeff}

In this section, we derive the Navier-Stokes equation and the corresponding transport coefficients for the interacting QGP system described by the set of Boltzmann equation given in Eqs.~\eqref{Beq_Q_fin}-\eqref{Beq_G_fin}. We also calculate the individual contribution to the transport coefficients from quarks, anti-quarks and gluons.

\subsection{Non-equilibrium corrections to distribution functions}

In order to derive the transport coefficients corresponding to various dissipative quantities, we evaluate the integrals in Eqs.~\eqref{pimunu_KT}--\eqref{nmu_KT}. The first step towards evaluating these integrals is to obtain the out-of-equilibrium parts of the distribution functions, $\delta Q$, $\delta\bar{Q}$, and $\delta G$. We use Eq.~\eqref{Beq_F} to derive these out-of-equilibrium parts in a compact form. For distributions which are slightly away from equilibrium, we have $\textbf{F} = \textbf{F}_{\rm eq} + \delta \textbf{F}$, where
\begin{align}\label{delta F}
\delta \textbf{F}(x,p) = 
\begin{bmatrix}
\,\delta Q(x,p)\,\\
\,\delta\bar{Q}(x,p)\,\\
\,\delta G(x,p)\,
\end{bmatrix}.
\end{align}
Keeping in mind that $\hat{R}_{\rm tr} \textbf{F}_{\rm eq} = 0$, Eq.~\eqref{Beq_F} can be rewritten as
\begin{align}\label{Beq_F_delF}
p^\mu\partial_\mu \textbf{F} = -(u\!\cdot\! p)\,\hat{R}\,\delta\textbf{F},
\end{align}
where $\hat{R} = \hat{R}_{\rm eq} + \hat{R}_{\rm tr}$ and is given by the following matrix
\begin{align}\label{R}
\hat{R} = \frac{1}{2}\!
\begin{bmatrix}
\,2 \nu_{\rm eq} + \nu^{qg}_{\rm tr}\, &  \nu^{qg}_{\rm tr} & -  \nu^{gq}_{\rm tr}\\
\nu^{qg}_{\rm tr} & \,2 \nu_{\rm eq} +  \nu^{qg}_{\rm tr}\, & -  \nu^{gq}_{\rm tr}\\
-2 r \nu^{qg}_{\rm tr} & - 2 r \nu^{qg}_{\rm tr} & \,2 \nu_{\rm eq} + 2 r \nu^{gq}_{\rm tr}\,
\end{bmatrix}.
\end{align}
In the above equation, we have used the notations given in Eq.~\eqref{tau_nu}.

Using Eq.~\eqref{Beq_F_delF}, we obtain the first-order gradient correction to the vector distribution function,
\begin{align}\label{deltaF1}
\delta\textbf{F} = -\frac{1}{(u\!\cdot\! p)}\hat{R}^{-1} p^\mu \partial_\mu \textbf{F}_{\rm eq},
\end{align}
where $\hat{R}^{-1}$ is the inverse of the matrix $\hat{R}$. This inverse matrix is given by
\begin{align}\label{R_inv}
\hat{R}^{-1} = \tau_{\rm eq}\!
\begin{bmatrix}
    A & -B & C\\
    -B & A & C\\
    D & D & E
\end{bmatrix}
\end{align}
where $A\!=\!1\!-\![2\!+\!4 \cosh{\alpha} (\gamma \sinh{\alpha}\!+\!r)]^{-1}$, $B\!=\!1\!-\!A$, $C=2B\cosh\alpha$, $D=2Br$, $E=1-2Cr$. 

From Eqs.~\eqref{deltaF1}--\eqref{R_inv}, we finally obtain the non-equilibrium corrections to the distribution functions 
\begin{align}
\delta Q &=\frac{\tau_{\rm eq}}{(u\!\cdot\! p)}\!\Big[ B\, p^\mu \partial_\mu \bar{Q}_{\rm eq} \!-\! A\, p^\mu \partial_\mu Q_{\rm eq} \!- C p^\mu \partial_\mu G_{\rm eq} \Big]\!, \label{delQ}\\
\delta\bar{Q}&=\frac{\tau_{\rm eq}}{(u\!\cdot\! p)}\!\Big[ B\, p^\mu \partial_\mu Q_{\rm eq} \!-\! A\, p^\mu \partial_\mu \bar{Q}_{\rm eq} \!- C p^\mu \partial_\mu G_{\rm eq} \Big]\!, \label{delQbar}\\
\delta G &= \frac{-\tau_{\rm eq}}{(u\!\cdot\! p)}\!\Big[ D p^\mu \partial_\mu Q_{\rm eq} \!+\! D p^\mu \partial_\mu \bar{Q}_{\rm eq} \!+\! E p^\mu \partial_\mu G_{\rm eq} \Big]. \label{delG}
\end{align}

\subsection{Navier-Stokes relations and transport coefficients} 

The derivatives of the equilibrium distribution functions in the above equations lead to derivatives of $u^\mu$, $\beta$ and $\alpha$. Some of these derivatives can be eliminated using first-order hydrodynamic equations. In particular, using the equations for hydrodynamic evolution, Eqs.~\eqref{continuity}--\eqref{diffusion}, along with the expressions for the thermodynamic quantities, Eqs.~\eqref{energy_density}--\eqref{number_density}, we obtain
\begin{equation}\label{deriv_alpha_beta}
\dot{\alpha} = \chi_\alpha\,\theta, ~~ \dot{\beta} = \chi_\beta\,\theta, ~~
\nabla^\mu \beta = -\beta\, \dot{u}^\mu + \frac{n}{\epsilon + P} \nabla^\mu \alpha, 
\end{equation}
where
\begin{align}
\chi_\alpha &\equiv \frac{(2g_q \cosh{\alpha}\!+\!g_g) I_{30}\, n - 2g_q \sinh{\alpha}\, I_{20} (\epsilon \!+\! P)}{2\, g_q\,D_{20}}, \label{chia}\\
\chi_\beta &\equiv \frac{\sinh{\alpha}\,I_{20}~n - \cosh{\alpha}\,I_{10}~(\epsilon + P)}{D_{20}}. \label{dot beta}
\end{align}
Here the thermodynamic integrals and the coefficient $D_{20}$ are defined as
\begin{align}
I_{nq} &\equiv \frac{1}{(2q + 1)!!}\int dp \left( u\!\cdot\! p \right)^{n - 2q} \left( \Delta_{\alpha\beta}\, p^\alpha p^\beta \right)^q G_{\rm eq}, \label{Inq}\\
D_{20} &\equiv 2 g_q \sinh^2\!\alpha\,I_{20}^2 \!- (2 g_q \cosh{\alpha} + g_g)\cosh{\alpha}\,I_{30} I_{10}. \label{D20}
\end{align}
These relations can be used to derive the relativistic Navier-Stokes equations connecting the shear stress tensor, bulk pressure, and baryon charge conductivity with the shear flow tensor, expansion scalar, and transverse gradient of $\alpha$.

Substituting Eqs.~\eqref{delQ}--\eqref{delG} into Eqs.~\eqref{pimunu_KT}--\eqref{nmu_KT}, using Eqs.~\eqref{deriv_alpha_beta}--\eqref{D20}, and performing the integrals, we obtain 
\begin{equation}
\pi^{\mu\nu} = 2 \eta\, \sigma^{\mu\nu}, \quad \Pi = - \zeta\, \theta, \quad n^\mu = \kappa\, (\nabla^\mu \alpha), \label{NS}
\end{equation}
where $\eta$, $\zeta$ and $\kappa$ are the coefficient of shear viscosity, coefficient of bulk viscosity and baryon charge conductivity, respectively. These transport coefficients are given by the following expressions
\begin{align}
\frac{\eta}{\tau_{\rm eq}} &= \left( g_g + 2\, g_q \cosh\alpha \right) \beta\, I_{32}, \label{eta}\\ 
\frac{\zeta}{\tau_{\rm eq}} &= \left(\frac{5}{3} + \frac{\chi_\beta\, I_{31}}{\beta\, I_{32}} \right)\!\frac{\eta}{\tau_{\rm eq}} +  \chi_\alpha\, n\, T,
\label{zeta} \\
\frac{\kappa}{\tau_{\rm eq}}  &= - \frac{ n^2\, T}{\epsilon + P} - 2\, g_q\, I_{11} \cosh{\alpha} . \label{kappa} 
\end{align}
From the above equations, we see that the transport coefficients $\eta$, $\zeta$ and $\kappa$ do not depend on $\gamma$ and hence on the transition time scale. This can be understood from Eqs.~\eqref{Tmunu} and \eqref{Nmu} where we see that the energy-momentum tensor and net particle current depend on $S^+$ and $Q^-$, respectively. However, the evolution of $Q^-$ and $S^+$ does not depend on $\tau_{\rm tr}$ as demonstrated in Eq.~\eqref{eq:Qminuseq} and \eqref{eq:Spluseq}. Also, note that despite an overall negative sign in Eq.~\eqref{kappa}, $\kappa$ is positive because $I_{11}$ is a negative quantity; see Eq.~\eqref{I11}.

\subsection{Quark, antiquark and gluon contributions to transport coefficients} 

Splitting the total energy-momentum tensor into individual components describing quarks, antiquarks, and gluons, we can identify the viscosity coefficients characterizing different components of the system. In particular, we obtain the three shear viscosity coefficients:
\begin{align}
\frac{\eta_Q}{\tau_{\rm eq}} &=   g_q e^\alpha \beta\, I_{32}, \label{etaq}\\ 
\frac{\eta_{\bar Q}}{\tau_{\rm eq}} &=   g_q e^{-\alpha} \beta\, I_{32}, \label{etaaq}\\ 
\frac{\eta_G}{\tau_{\rm eq}} &=   g_g  \beta\, I_{32}. \label{etag}
\end{align}
Equations \eqref{etaq}--\eqref{etag} show that the partial shear viscosities are independent of the transition time. The total shear viscosity is given by the sum $\eta=\eta_Q+\eta_{\bar Q}+\eta_G$ and turns out to be positive. Similarly, we can write $\zeta=\zeta_Q+\zeta_{\bar Q}+\zeta_G$, where
\begin{align}
\frac{\zeta_{Q}}{\tau_{\rm eq}} &= \left(\frac{5}{3} \!+\! \frac{\chi_{\beta}\, I_{31}}{\beta\, I_{32}}\!\right)\! \frac{\eta_{Q}}{\tau_{\rm eq}} \!- g_q \!\left( A e^{\alpha} \!+\! B e^{-\alpha}\right) \chi_{\alpha}\, I_{21}, \label{zeta_Q} \\
\frac{\zeta_{\bar{Q}}}{\tau_{\rm eq}} &= \left(\frac{5}{3} \!+\! \frac{\chi_{\beta}\, I_{31}}{\beta\, I_{32}}\right)\! \frac{\eta_{\bar{Q}}}{\tau_{\rm eq}} \!+ g_q \!\left( B e^{\alpha} \!+\! A e^{-\alpha}\right) \chi_{\alpha}\, I_{21}, \label{zeta_Qbar} \\
\frac{\zeta_{G}}{\tau_{\rm eq}} &= \left(\frac{5}{3} \!+\! \frac{\chi_{\beta}\, I_{31}}{\beta\, I_{32}}\right)\! \frac{\eta_{G}}{\tau_{\rm eq}} \!+ \frac{D\, \chi_{\alpha}}{r} n T. \label{zeta_G}
\end{align}
In this case, the partial contributions do depend on the transition time. It is interesting to note that the dependence of $\zeta_G$ on $\alpha$ is due to finite $\gamma$ and this dependence vanishes for $\gamma\to\infty$. We can also define the net conductivity by the difference $\kappa=\kappa_Q-\kappa_{\bar Q}$ with
\begin{align}
\frac{\kappa_{Q}}{\tau_{\rm eq}} &= g_q \bigg[ \left(\frac{n\, e^{\alpha}}{\epsilon + P}\right)\, I_{21} - \Big( A e^{\alpha} + B e^{-\alpha}\Big)\, I_{11} \bigg], \label{kappa_Q} \\
\frac{\kappa_{\bar{Q}}}{\tau_{\rm eq}} &= g_q \bigg[ \left(\frac{n\, e^{-\alpha}}{\epsilon + P}\right) I_{21} + \Big( B e^{\alpha} + A e^{-\alpha}\Big) I_{11} \bigg]. \label{kappa_Qbar}
\end{align}
Here, $\kappa_{Q}$ and $\kappa_{\bar{Q}}$ also depend on the transition time. 

\subsection{Limiting cases} 

It is interesting to separately consider the limit of zero baryon chemical potential. For $\alpha\to0$ (very low values of the $\mu/T$ ratio) and with other parameters fixed, including $\gamma$, one finds:
\begin{eqnarray}
\lim_{\alpha\to 0}\frac{\eta}{\tau_{\rm eq}} &=&  \left( g_g + 2\,g_q \right)  \beta I_{32},  \label{eq:alpha01} \\
\lim_{\alpha\to 0}\frac{\kappa}{\tau_{\rm eq}} &=& - 2\, g_q  I_{11},
\label{eq:alpha02} \\
\lim_{\alpha\to 0}\frac{\zeta}{\tau_{\rm eq}} &=&  \left( \frac{5}{3} - \frac{I_{31}^2}{I_{30}I_{32}}\right)\frac{\eta}{\tau_{\rm eq}}\bigg|_{\alpha\to 0}.
\label{eq:alpha03}
\end{eqnarray}
We note that for low values of $z$, one can use the expansions of the thermodynamic integrals given by Eqs.~\eqref{betaI32_z}-\eqref{eq:usedalpha0}. The limits for individual contributions are as follows:
\begin{eqnarray}
\lim_{\alpha\to 0}\frac{\eta_Q}{\eta}&=&\lim_{\alpha\to 0}\frac{\zeta_Q}{\zeta}  =   \frac{r}{1+2r}, \\
\lim_{\alpha\to 0}\frac{\eta_{\bar Q}}{\eta} &=&\lim_{\alpha\to 0}\frac{\zeta_{\bar Q}}{\zeta}=   \frac{r}{1+2r},  \\
\lim_{\alpha\to 0}\frac{\eta^G}{\eta} &=& \lim_{\alpha\to 0}\frac{\zeta^G}{\zeta} =  \frac{1}{1+2r}, \\
\lim_{\alpha\to 0}\frac{\kappa^Q}{\kappa} &=& -\lim_{\alpha\to 0}\frac{\kappa^{\bar{Q}}}{\kappa} =  \frac{1}{2}, \label{eq:limits}
\end{eqnarray}

\begin{figure}[t]
\begin{center}
\includegraphics[width=8.5cm]{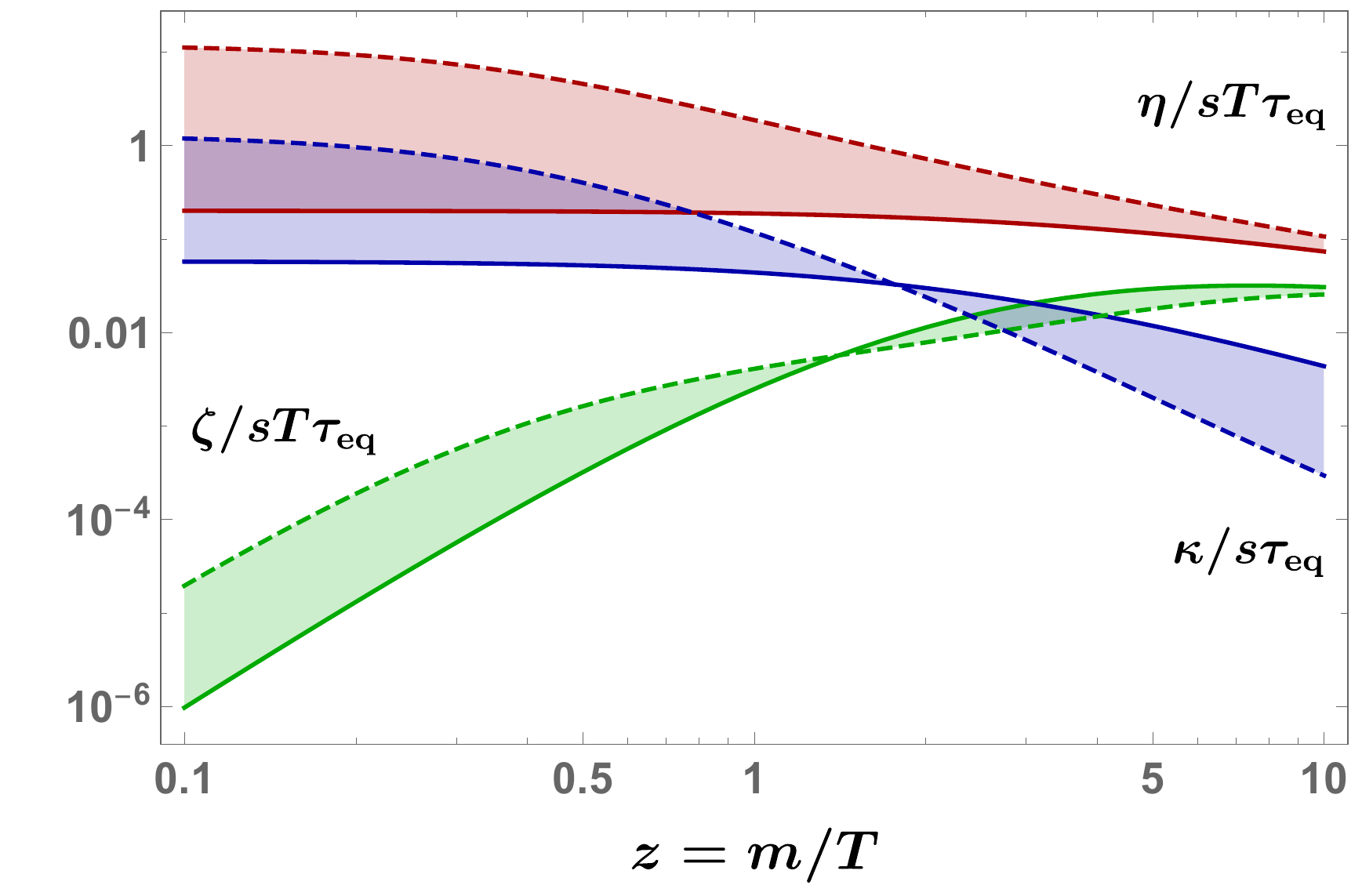}
\end{center}
\caption{\small 
The $m/T$ dependence of the kinetic coefficients $\eta$, $\zeta$, and $\kappa$, rescaled by $s T \tau_{\rm eq}$ (shear and bulk viscosities) and $s \tau_{\rm eq}$ (baryon conductivity). The parameter
$\alpha$ ranges from $0.1$ (solid lines) to $4$ (dashed lines).
}
\label{fig:all}
\end{figure}


\section{Results and Discussions}

\begin{figure}[t]
\begin{center}
\includegraphics[width=8.5cm]{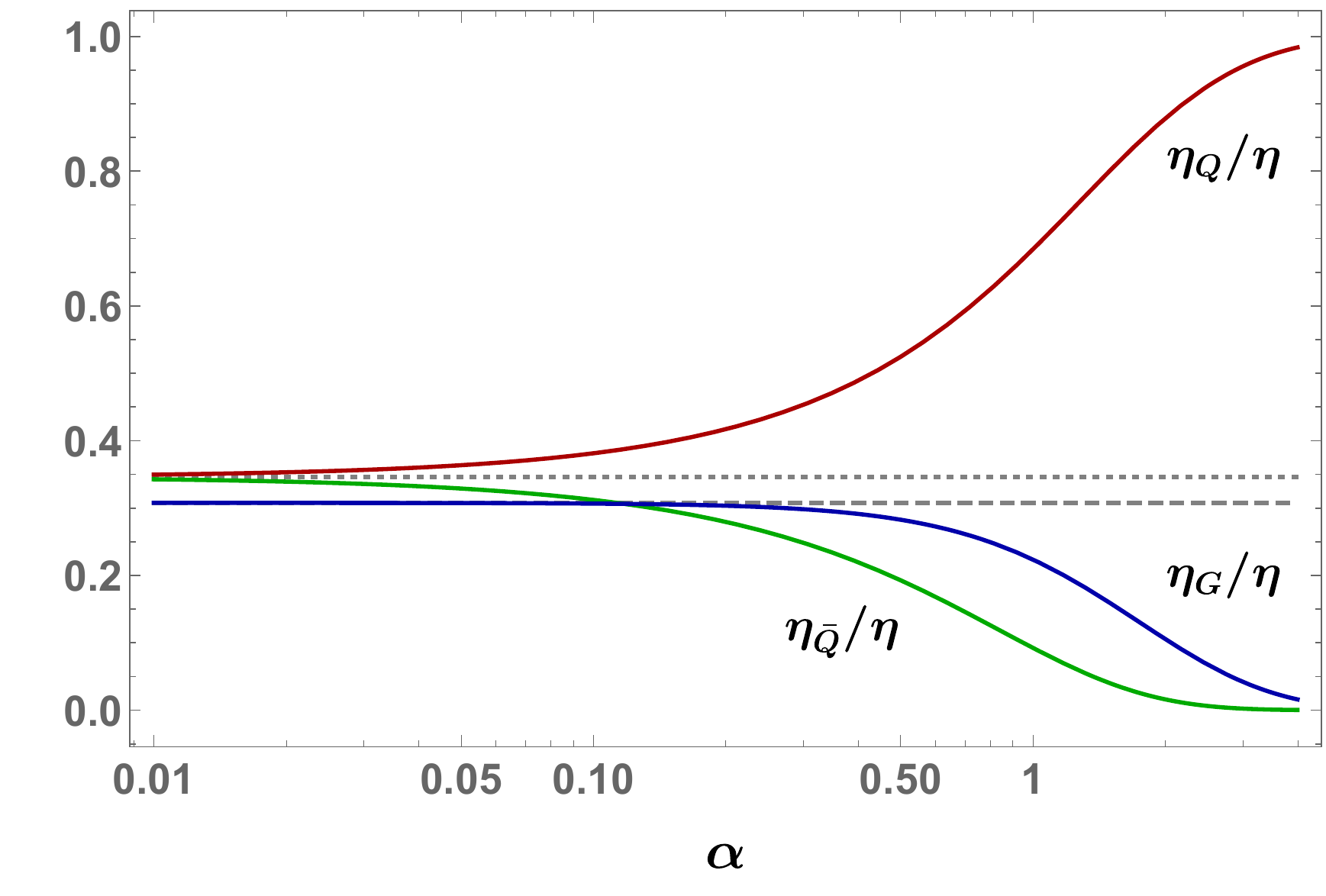}
\end{center}
\caption{\small The individual contributions to the shear viscosity coefficient, rescaled by the total value of $\eta$ and shown as functions of $\alpha$. In this case, the presented ratios are functions of $\alpha$ and $r$ only.  }
\label{fig:eta}
\end{figure}

In the previous section, we showed that the coefficient of shear viscosity, $\eta$, is independent of $\gamma$ and hence the transition time-scale. In this section we present our numerical results describing the two viscosity coefficients and the baryon conductivity based on Eqs.~\eqref{eta}--\eqref{kappa}. We also evaluate numerically the individual contribution from quarks, anti-quarks and gluons to the coefficient of bulk viscosity, $\zeta$ and the charge conductivity, $\kappa$, based on Eqs.~\eqref{zeta_Q}-\eqref{kappa_Qbar}.

\begin{figure}[t]
\begin{center}
\includegraphics[width=8.5cm]{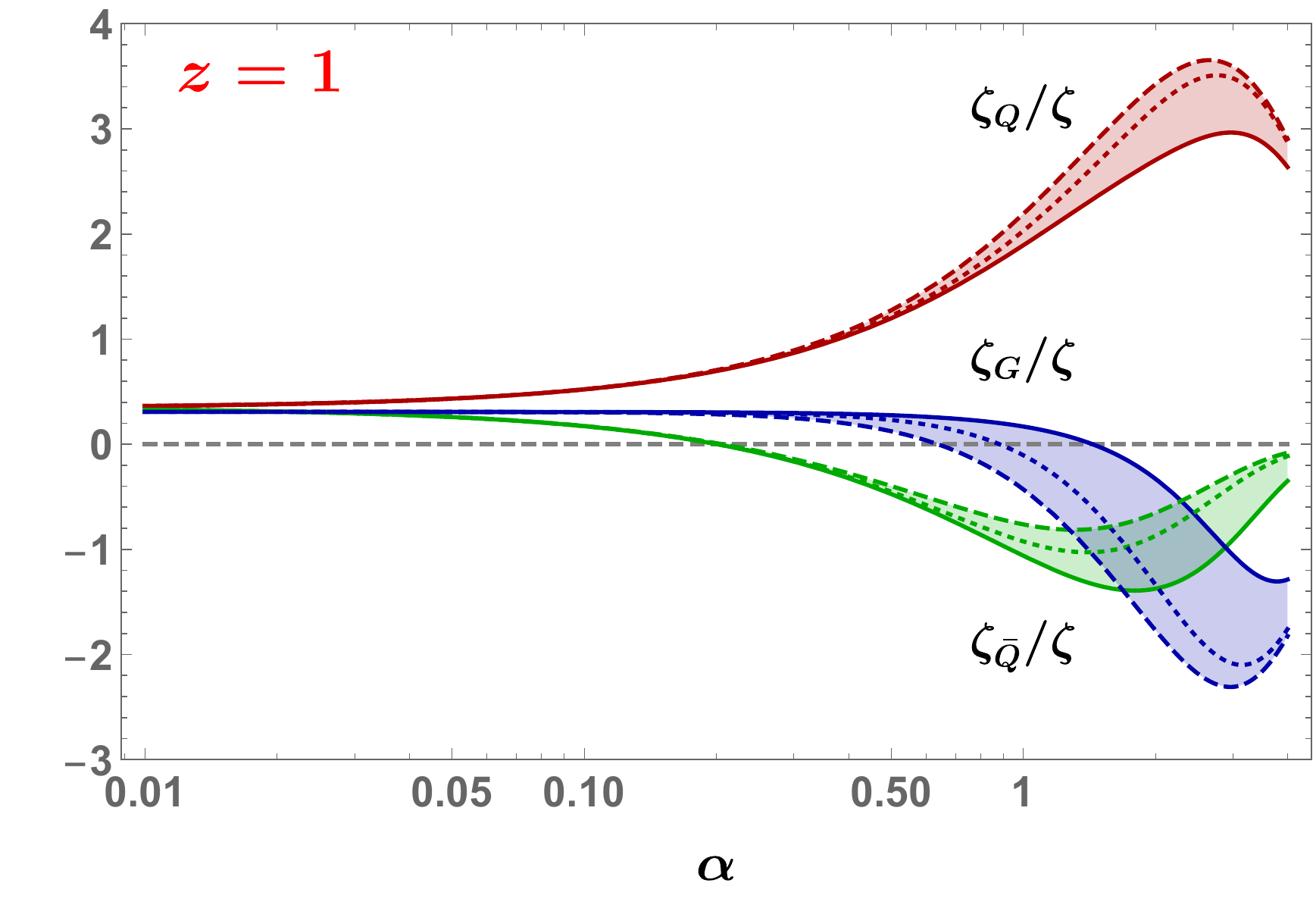}
\end{center}
\caption{\small The individual contributions to the bulk viscosity coefficient, rescaled by the total value of $\zeta$ and calculated for $z=1$. The parameter $\gamma$ varies from $0.1$ (solid lines), through $1$ (dotted lines) to $10$ (dashed lines). Note that in certain regions of the parameter space $\zeta_G$ and $\zeta_{\bar Q}$ become negative. }
\label{fig:zetagamma}
\end{figure}

Figure~\ref{fig:all} shows the dependence of the three considered kinetic coefficients on the ratio $z=m/T$. The values of the shear and bulk viscosities are rescaled by the factor $s T \tau_{\rm eq}$, while the baryon conductivity is rescaled by $s \tau_{\rm eq}$. At low temperatures and small values of the chemical potential, the rescaled value of the shear viscosity approaches the value of 1/5 \cite{Florkowski:2013lya}. For small masses, i.e. for $z < 1$, the bulk viscosity is significantly smaller compared to the shear viscosity. We find that all the kinetic coefficients for the mixture are positive, as required by the second law of thermodynamics. 

\begin{figure}[t]
\begin{center}
\includegraphics[width=8.5cm]{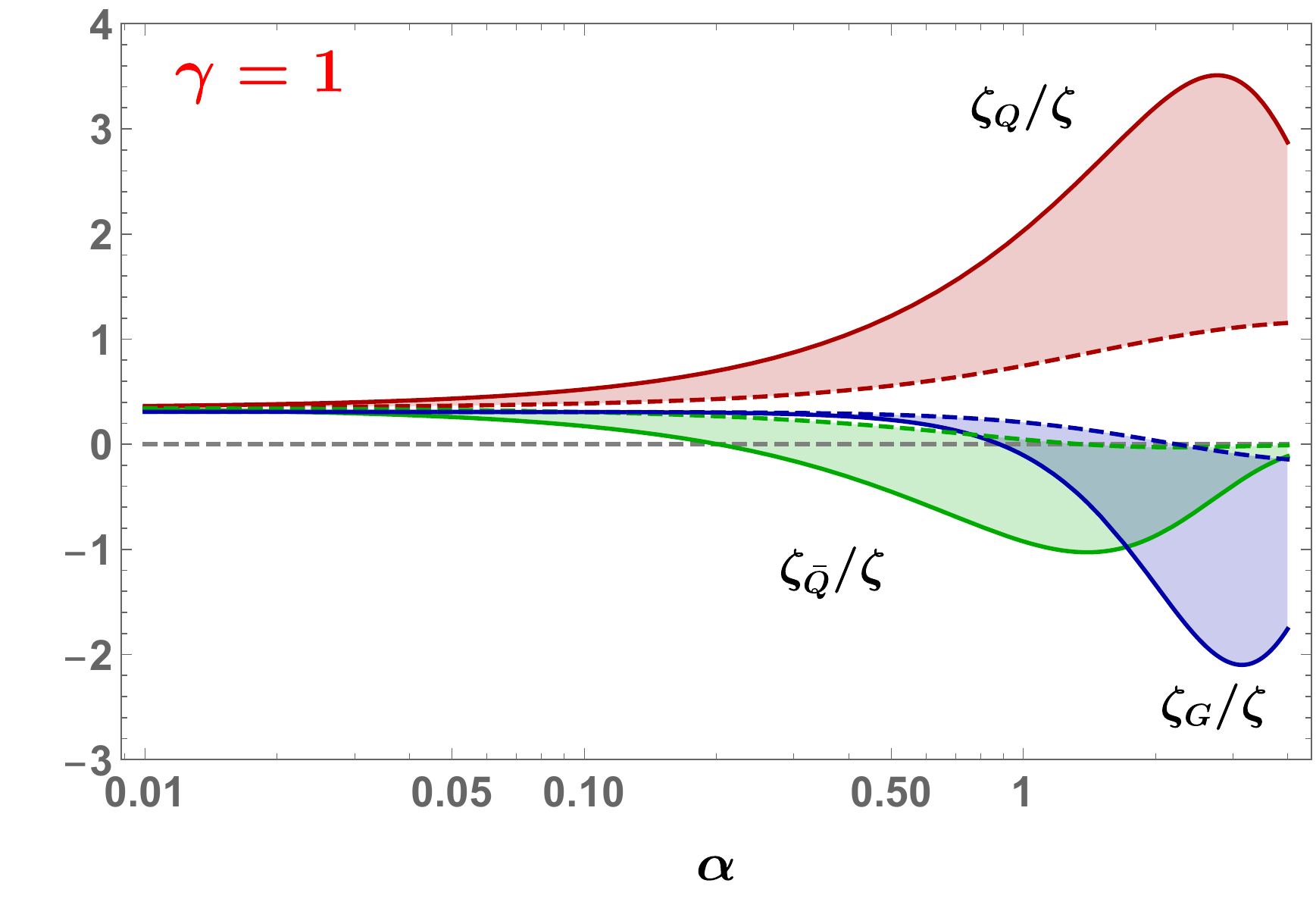}
\end{center}
\caption{\small Same as Fig.~\ref{fig:zetagamma} but for $\gamma=1$. The ratio $z$ varies from $1$ (solid lines), to $10$ (dashed lines).}
\label{fig:zetaz}
\end{figure}

In Fig.~\ref{fig:eta} we show individual contributions to the shear viscosity, plotted now as functions of $\alpha$. In view of the arguments presented below Eq.~\eqref{G_eq} we restrict our calculations to the range $0 \leq \alpha < 4$. As expected, for very large values of $\alpha$ the shear viscosity is dominated by the quark contribution. At low values of $\alpha$, the individual contributions are given by the fractions involving internal degeneracy factors, see Eq.~\eqref{eq:limits}. In the numerical calculations we use $r=9/8$, note that for $r=1$ all the contributions would be the same and equal to 1/3.

\begin{figure}[t]
\begin{center}
\includegraphics[width=8.5cm]{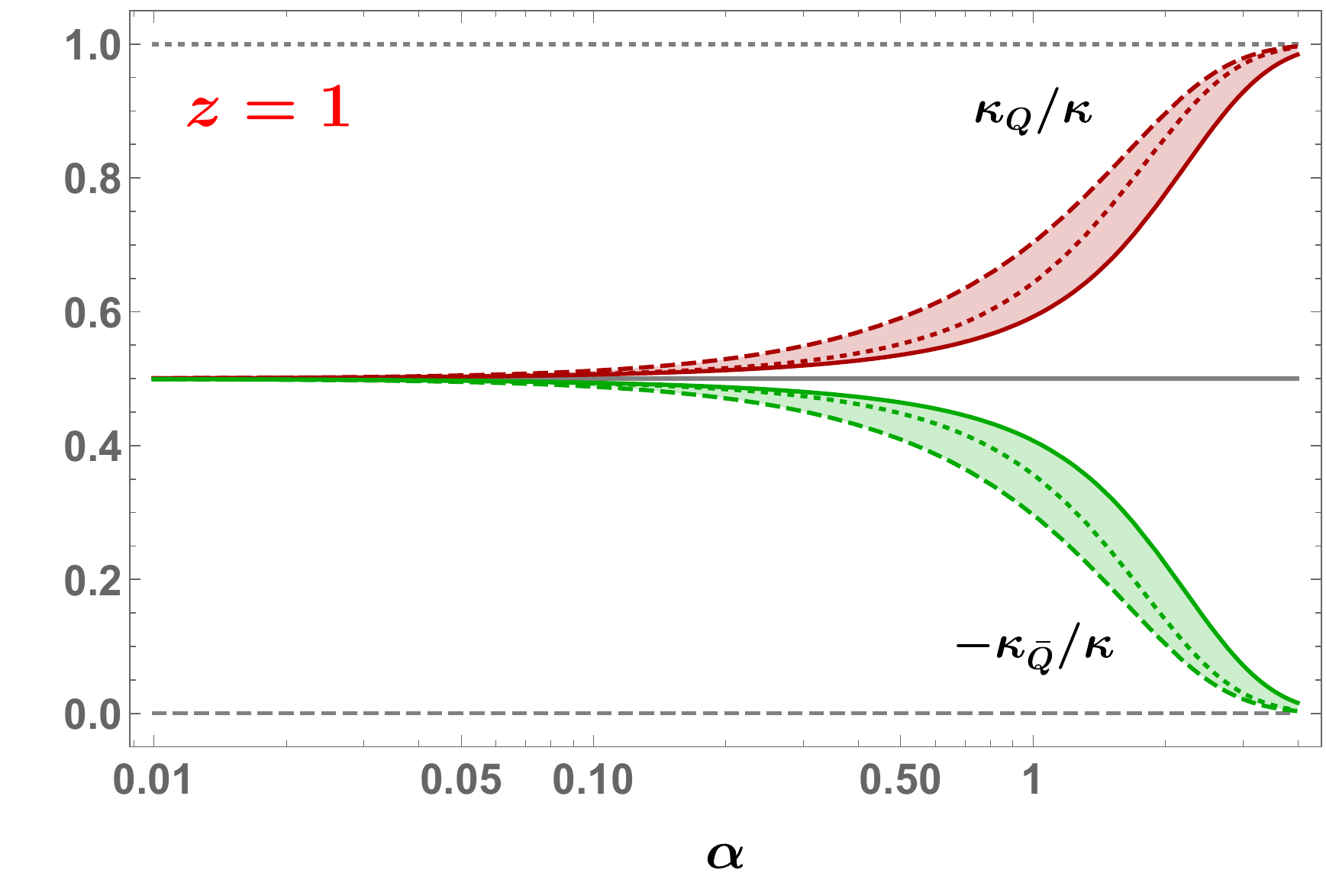}
\end{center}
\caption{\small The two contributions to the baryon conductivity coefficient for $z=1$, $\gamma$ varies from $0.1$ (solid lines), through $1$ (dotted lines) to $10$ (dashed lines). }
\label{fig:kappagamma}
\end{figure}

Figures (\ref{fig:zetagamma}) and (\ref{fig:zetaz}) show the $\alpha$ dependence of the individual contributions to the bulk viscosity coefficient $\zeta$. We find a non-trivial dependence on the transition rate quantified by the value of $\gamma$, Fig.~(\ref{fig:zetagamma}), and on the mass over temperature ratio $z$, Fig.~(\ref{fig:zetaz}). Interestingly, in some regions of the parameter space the individual contributions turn out to be negative. Nevertheless, the total value of the bulk viscosity is always positive as demonstrated in Fig.~1.

\begin{figure}[t]
\begin{center}
\includegraphics[width=8.5cm]{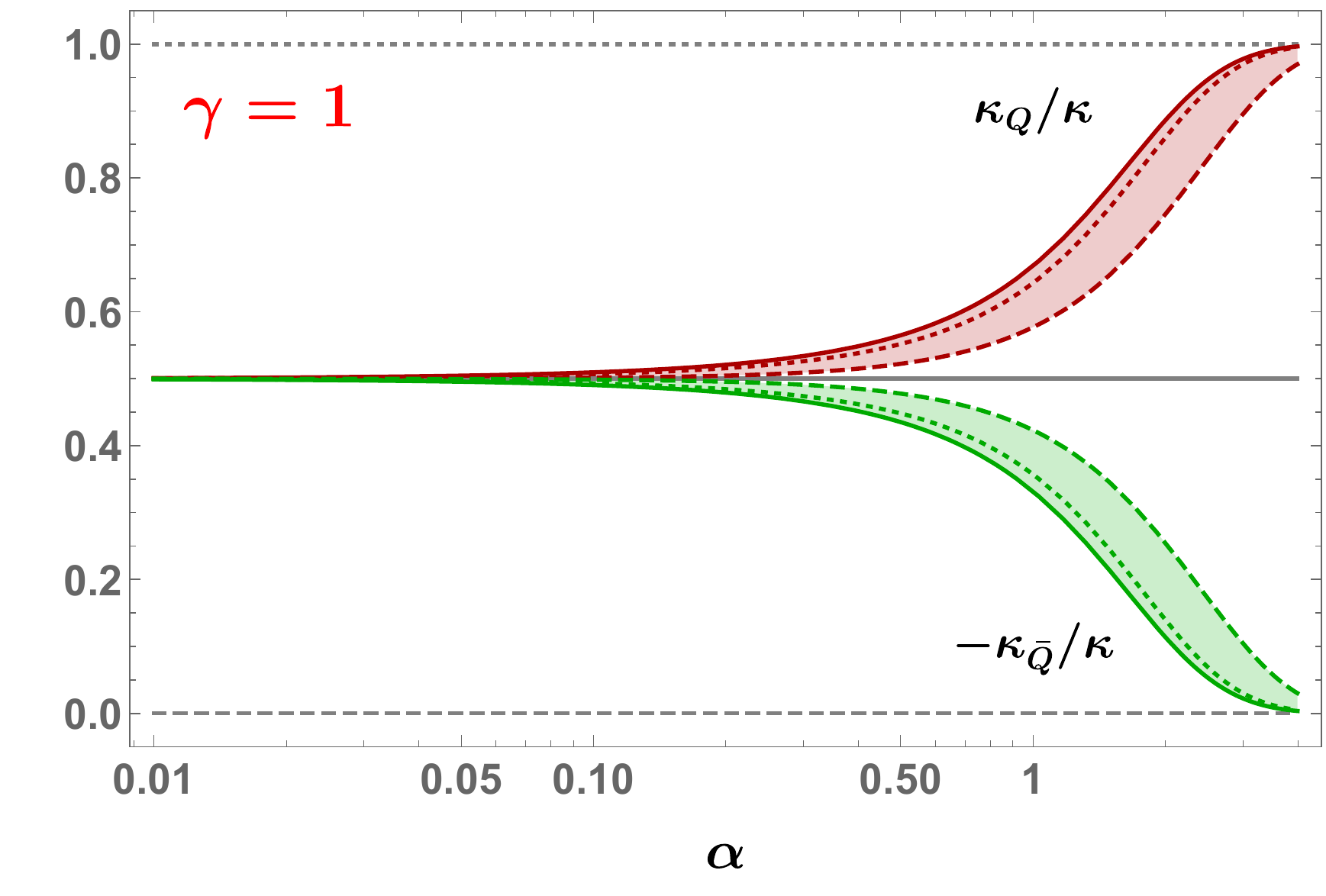}
\end{center}
\caption{\small The two contributions to the baryon conductivity coefficient for $\gamma=1$, $z$ varies from $0.1$ (solid lines), through $1$ (dotted lines) to $10$ (dashed lines). }
\label{fig:kappaz}
\end{figure}

Finally, in the similar way in Figs.~(\ref{fig:kappagamma}) and (\ref{fig:kappaz}) we show the $\alpha$ dependence of the two contributions to the baryon conductivity. Here again, one can find a non-trivial dependence on the transition time and the $m/T$ ratio.


\section{Summary and Outlook}

In this work, we have extended the Boltzmann equation in the relaxation time approximation (RTA) to include explicitly the transitions between quarks and gluons. We considered inelastic interactions for the quark antiquark annihilation to gluon and corresponding pair production. Using the detailed balance condition as well as conditions of energy-momentum and current conservation, obtained from the Boltzmann equations, we demonstrated that there exists only two independent relaxation time scales in such an interacting system. Subsequently we derived first-order dissipative hydrodynamic equations for the evolution of this interacting system and obtained their corresponding transport coefficients. We found that the detailed balance condition renders the transport coefficients of the plasma insensitive to the newly proposed transition time-scale. On the other hand, we showed that the individual contributions due to quarks, anti-quarks and gluons are dependent on this new time-scale.

It is interesting to note that the out-of-equilibrium parts of the distribution functions, given by Eqs.~\eqref{delQ}-\eqref{delG}, also depend on this new time-scale. These contributions are important for thermal particle production in the medium. Therefore, electromagnetic probes such as photons and dileptons, which are sensitive to the evolution of the QGP, are expected to be affected by this transition time-scale. Looking forward, it will be interesting to estimate the dependence of these probes on the transition time-scale and this is left for future works. Moreover, the present method of generalizing the RTA Boltzmann equation, to explicitly include inelastic collisions, can be applied to other processes as well. In that sense, we have presented a powerful framework to model different processes within RTA. It will be interesting to consider different processes and construct a set of RTA Boltzmann equations. It will also be interesting to extend the current derivation to momentum dependent relaxation times and quantum statistics, as well as to obtain causal second-order dissipative equations. We leave these problems for future work.

\begin{acknowledgements}
S.B. and A.J. thank Jean-Yves Ollitrault, Sunil Jaiswal and Rajeev Bhalerao for useful discussions. W.F., A.J. and R.R. acknowledge kind hospitality of ExtreMe Matter Institute EMMI at GSI Darmstadt where this work was initiated. S.B. and A.J. acknowledge kind hospitality of Jagiellonian University and Institute of Nuclear Physics, Krakow, where part of this work was completed. A.J. was supported in part by the DST-INSPIRE faculty award under Grant No. DST/INSPIRE/04/2017/000038. W.F. and R.R. were supported in part by the Polish National Science Center Grants No. 2016/23/B/ST2/00717 and No. 2018/30/E/ST2/00432, respectively.
\end{acknowledgements}

\appendix

\section{Thermodynamic Integrals}

The thermodynamic integrals $I_{nq}$, which are frequently used in this work, can be expressed in terms of well-known special functions. Since the defining  integrals are Lorentz scalars, they can be computed in the local fluid rest frame defined by the condition $u^\mu=(1,\textbf{0})$.  In this way, we obtain the following expressions:
\begin{align}
I_{10} &= \frac{z^2 T^3}{2\pi^2} K_2(z), \label{I10}\\
I_{11} &= - \frac{z^3 T^3}{24\pi^2} \left[K_3(z) - 5 K_1(z) +  4 K_{i,1}(z)\right], \label{I11}\\
I_{20} &= \frac{z^2 T^4}{2\pi^2}\left[3 K_2(z) + z K_1(z)\right], \label{I20}\\
I_{21} &= - \frac{z^2 T^4}{2\pi^2} K_2(z), \label{I21}\\
I_{30} &= \frac{z^5 T^5}{32\pi^2}\left[K_5(z) + K_3(z) - 2 K_1(z)\right], \label{I30}\\
I_{31} &= - \frac{z^5 T^5}{96\pi^2}\left[K_5(z) - 3 K_3(z) + 2 K_1(z)\right], \label{I31}\\
I_{32} &= \frac{z^5 T^5}{480\pi^2}\left[K_5(z) - 7 K_3(z) + 22 K_1(z) - 16 K_{i,1}(z)\right], \label{I32}
\end{align}
where $K_n(z)$ denotes the modified Bessel functions of the second kind with the argument $z=m/T$. Here
\begin{align}
K_{i,1}(z) &\equiv \int_0^\infty d\theta\sech\theta\exp(-z\cosh\theta)\\
&=\frac{\pi}{2} \left[ 1 - z K_0(z) L_{-1}(z) - z K_1(z) L_0(z) \right]
\end{align}
is the first order Bickley-Naylor function  with $L_i$ being the modified Struve function.

The small-$z$ expansions of the thermodynamic integrals that can be applied to Eqs.~\eqref{eq:alpha01}-\eqref{eq:alpha03} are
\begin{eqnarray}
&& \beta I_{32} \approx \frac{4 T^4}{5 \pi^2}\left[ 1-\frac{5z^2}{24}+{\cal O }\!\left(z^3\right)\right], \label{betaI32_z}\\
&&  I_{11} \approx -\frac{ T^3}{3 \pi^2}\left[ 1-\frac{3z^2}{4}+{\cal O }\!\left(z^3\right)\right],  \label{I11_z} \\ 
&&  \frac{5}{3} - \frac{I_{31}^2}{I_{30}I_{32}}  \approx \frac{25}{432}\Big[ z^4+{\cal O }\!\left(z^5\right)\Big]. \label{eq:usedalpha0}
\end{eqnarray}
In terms of the integrals $I_{nq}$, one can also express the energy density, pressure, and particle number density as:
\begin{align}
\epsilon &= \left( 2\,g_q\cosh\alpha + g_g \right) I_{20}, \\
P &= -\left( 2\,g_q\cosh\alpha + g_g \right) I_{21}, \\
n &= 2\,g_q\sinh\alpha\, I_{10}.
\end{align}
It is important to point out that the above relations for $\epsilon$, $P$, and $n$ are valid because the masses of quarks and gluons are assumed to be the same. We can further express some other important thermodynamic integrals as combinations of the above thermodynamic quantities:
\begin{align}
I_{30} &= \frac{3 \epsilon + (3 + z^2) P}{\beta \left( 2\,g_q\cosh\alpha + g_g \right)}, \\
I_{31} &= - \frac{\left(\epsilon + P\right)}{\beta \left( 2\,g_q\cosh\alpha + g_g \right)}.
\end{align}
Using these expressions we can write:
\begin{widetext}
\begin{align}
\chi_\alpha &= \left(\frac{ 2\,g_q\cosh\alpha + g_g }{2\, g_q}\right) \frac{(2g_q \cosh{\alpha}\!+\!g_g) n \left[3 \epsilon + (3 + z^2) P\right] - 2\, \beta\, g_q\, \sinh{\alpha}\, \epsilon (\epsilon \!+\! P)}{2 \beta g_q\, \sinh^2\!\alpha\, \epsilon^2 \!- (2 g_q \cosh{\alpha} + g_g) \cosh{\alpha} \left[3 \epsilon + (3 + z^2) P\right] \beta P},   \\
\chi_\beta &= \beta \left( 2\,g_q\cosh\alpha + g_g\right) \frac{\sinh{\alpha}\, n\, \epsilon - \cosh{\alpha} (\epsilon + P) \beta P}{2 \beta g_q \sinh^2\!\alpha\, \epsilon^2 \!- (2 g_q \cosh{\alpha} + g_g)\cosh{\alpha} \left[3 \epsilon + (3 + z^2) P\right] \beta P}.
\end{align}

\end{widetext}

\bibliography{file1}

\end{document}